\documentclass{elsart}
\usepackage{graphicx}
\begin{document}
\begin{frontmatter}
\title{Quarkonium Spectral Function from Anisotropic Lattice}
\author{Alexander Velytsky}
\address{Department of Physics and Astronomy, UCLA, Los Angeles, CA 90095-1547, USA}
\begin{abstract}
We discuss the behavior of charmonia and bottomonia correlators and spectral functions 
above the deconfinement temperature and determine melting temperatures for
different mesonic states.
\end{abstract}

\end{frontmatter}

\section{Introduction}
\label{intro}
The study of heavy quarkonia at finite temperature is an important way to probe
hot and dense strongly interacting matter. 
General considerations suggest that quarkonia will melt at temperatures somewhat
higher than the deconfinement temperature as a result of modification of
interquark potential due to color screening. Matsui and Satz \cite{MS86} proposed that color screening at high
temperature will lead to quarkonium suppression, which can be used as a signal
of QGP formation. 
Studies of modification of quarkonium potential in terms of
spectral functions have started to appear only recently 
\cite{Datta:2002ck,umeda02,asakawa04,datta04}. However, the systematic errors were not well
controlled and it is not clear which features of the charmonium spectral
functions are physical and which are lattice artifacts of the MEM
analysis and/or finite lattice spacing.

This contribution reports updated results of our study of
charmonium \cite{Jakovac:2006bx,Petreczky:2005zy} and bottomonium
\cite{Petrov:2005ej} spectral functions and correlators at finite temperature.
The calculations are performed in quenched QCD using anisotropic lattices, the 
Wilson gauge action and the Fermilab fermion action \cite{fermilab}.
Further details of the lattice action and the parameters of the simulations
can be found in \cite{Jakovac:2006bx}.
Detailed results of the study and description of the Maximum Entropy Method (MEM)
will be presented in our upcoming paper \cite{jppv}.

\section{Meson Correlators and Spectral Functions}
We look at correlators of point meson operator
$J_H(t,x)=\bar q(t,x) \Gamma_H q(t,x)$,
where $\Gamma_H=1,\gamma_5, \gamma_{\mu}, \gamma_5 \gamma_{\mu}, 
\gamma_{\mu} \gamma_{\nu}$ fixes the mesonic channel to (pseudo)scalar,
(pseudo)vector and tensor channels. 
On the lattice one computes the Euclidean propagator 
$G_H(\tau,\vec{p})= \int
d^3x e^{i\vec{p}.\vec{x}}\langle T_{\tau} J_H(\tau, \vec{x}) J_H(0, \vec{0})
\rangle,$
which is related to the spectral function through the integral representation
\begin{equation}
G(\tau, \vec{p}) = \int_0^{\infty} d \omega
\sigma K(\omega, \tau), \quad
K(\omega, \tau) = \frac{\cosh(\omega(\tau-1/2
T))}{\sinh(\omega/2 T)}.\label{eq.kernel}
\nonumber
\end{equation}
Spectral functions can be reconstructed using Bayesian
techniques, such as the MEM \cite{asakawa01}. 
Instead of using Bryan's algorithm we use a new algorithm, which will be
described in details in \cite{jppv}.

\begin{figure}[ht]
\includegraphics[height=5cm,width=0.51\textwidth]{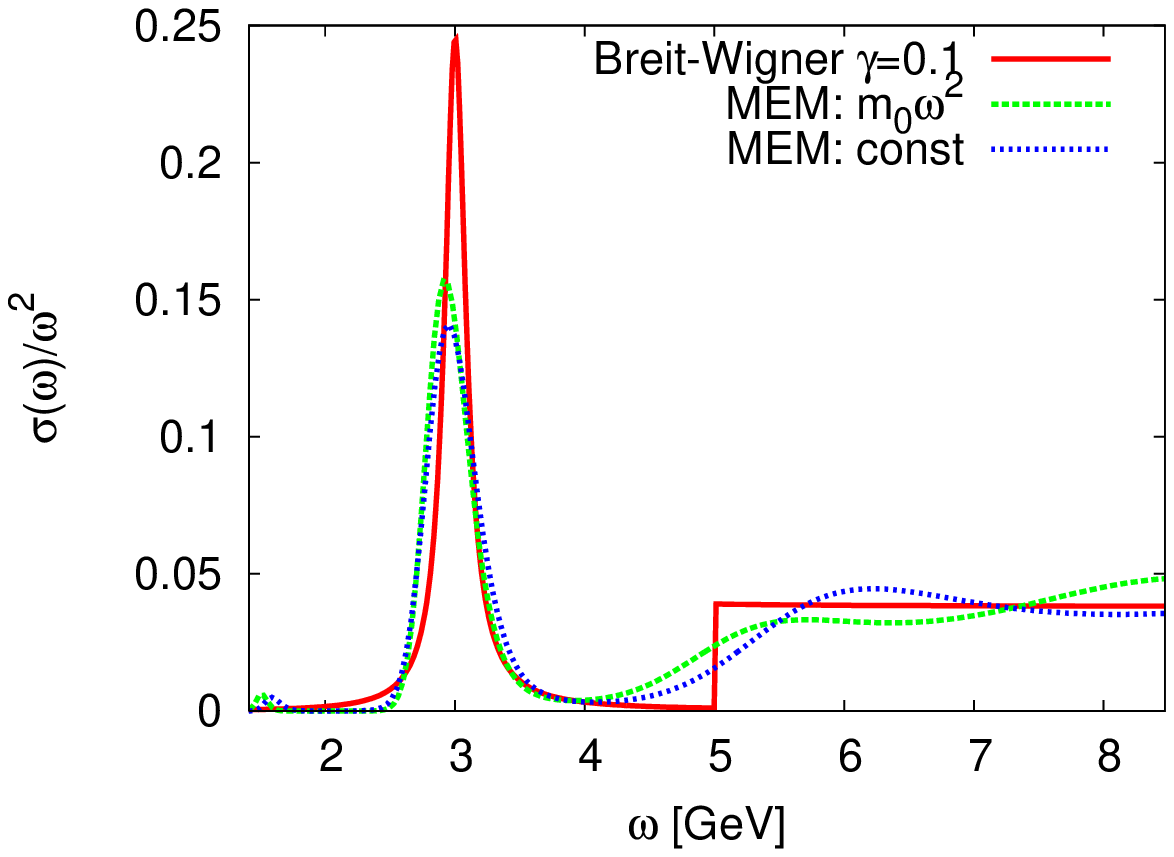}
\includegraphics[height=5cm,width=0.48\textwidth]{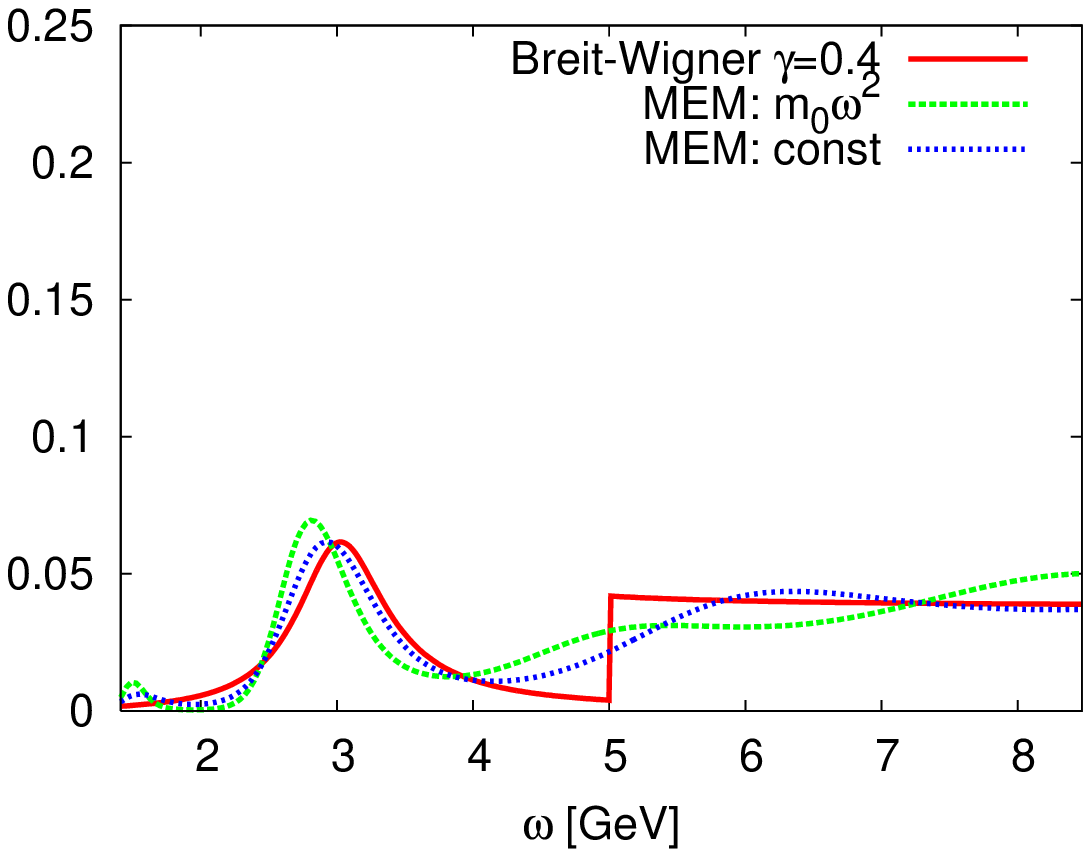}
\caption{Reconstruction of the $\gamma=0.1$ and $0.4$ GeV Breit-Wigner model
spectral functions.\label{fig:bw2}}
\end{figure}

First we check the MEM ability to reconstruct a model spectral
function of finite width. For a model spectral function we choose the
Breit-Wigner model spectral function, which (in physical units) is given by
\begin{equation}
\sigma(\omega)=r_1\frac 1\pi 
\frac{\gamma}
{(\omega-m)^2+\gamma^2}
\frac{\omega^3}{m^3}*\frac3{2\pi}.
\end{equation}
The
covariance matrix from an actual lattice simulation of charmonium at zero
temperature ($\beta=6.5$, $\xi=4$, $N_t=160$) is used to introduce the noise.
In Fig. \ref{fig:bw2} we show the model and reconstructed (with $m(\omega)=0.038\omega^2$
and $m(\omega)=1$ default models) spectral functions for the 
pseudoscalar channel. We see that
the MEM can successfully reconstruct the model spectral function. However, it is gradually
loosing
its ability as we increase the width and at $\gamma=2$ GeV is incapable of
spectral function reconstruction.

\section{Spectral functions and correlators at finite temperature}
\begin{figure}[ht]
\includegraphics[width=0.49\textwidth]{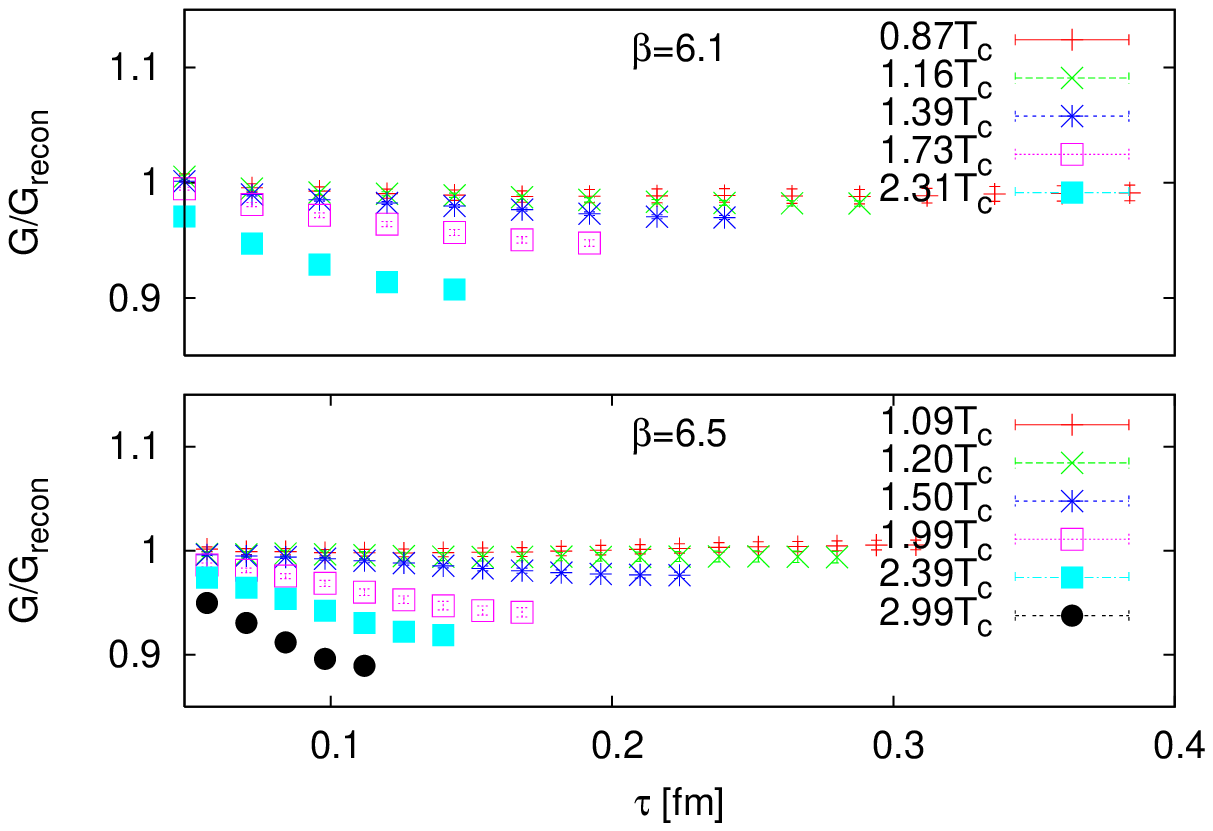}
\includegraphics[width=0.49\textwidth]{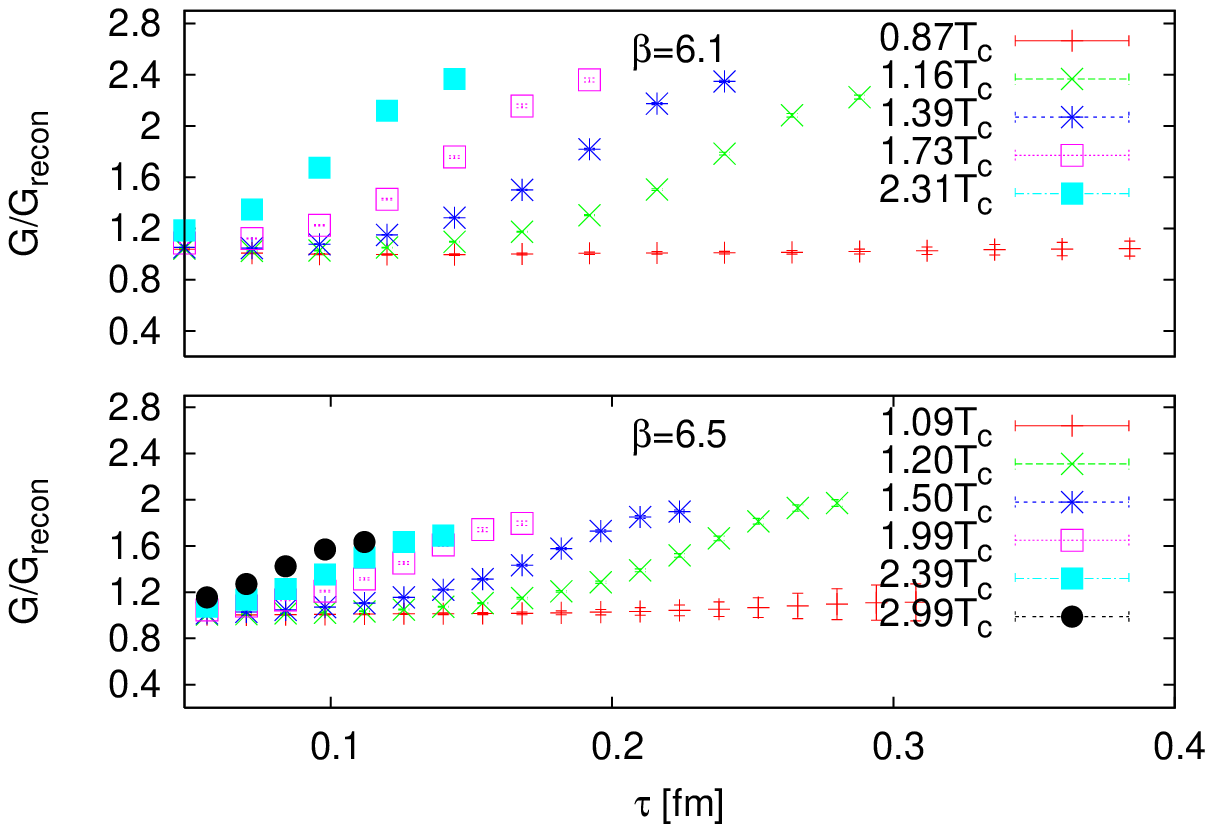}
\caption{The ratio $G(\tau,T)/G_{recon}(\tau,T)$ of charmonium for pseudoscalar
(left) and scalar (right) channels at $a_t^{-2}=8.18$ and $14.11$GeV
at different temperatures.}
\label{ratioc} 
\end{figure}

\begin{figure}[ht] 
\includegraphics[width=0.48\textwidth]{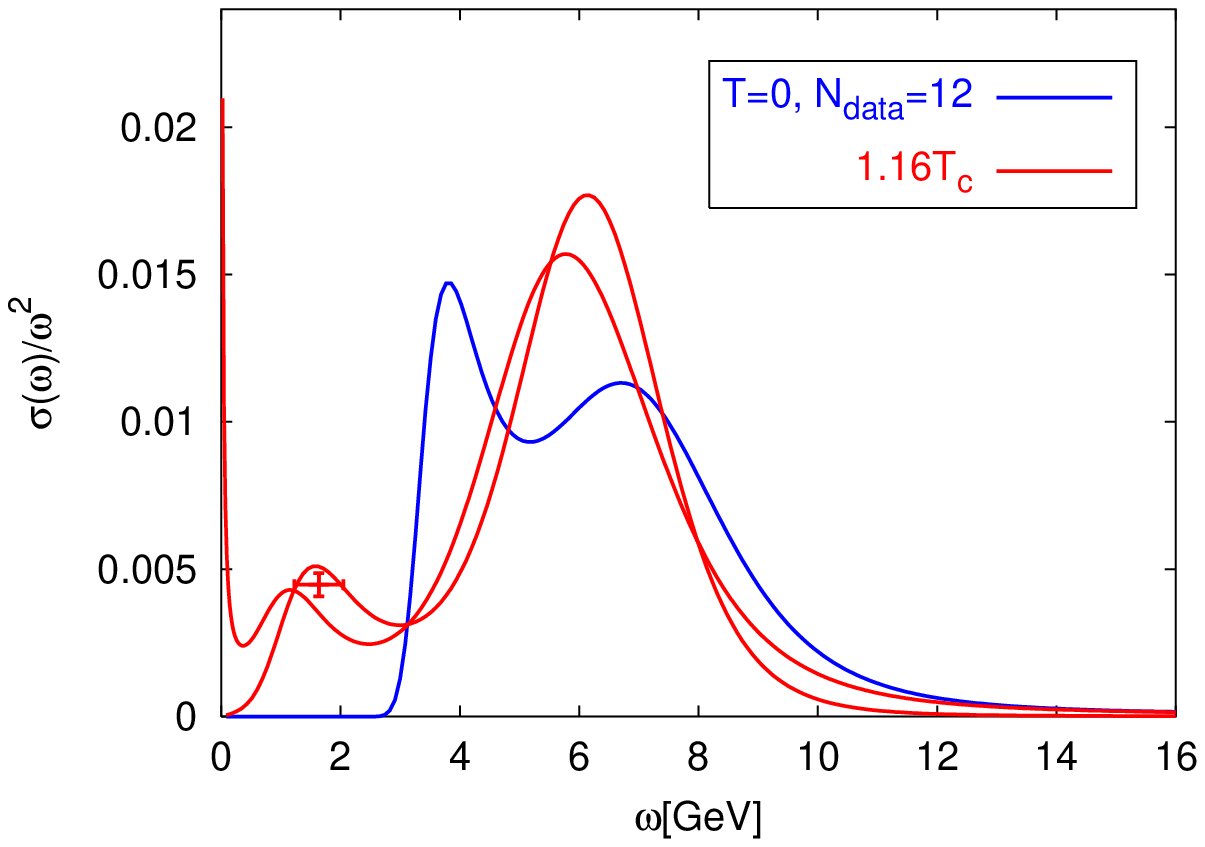}
\includegraphics[width=0.48\textwidth]{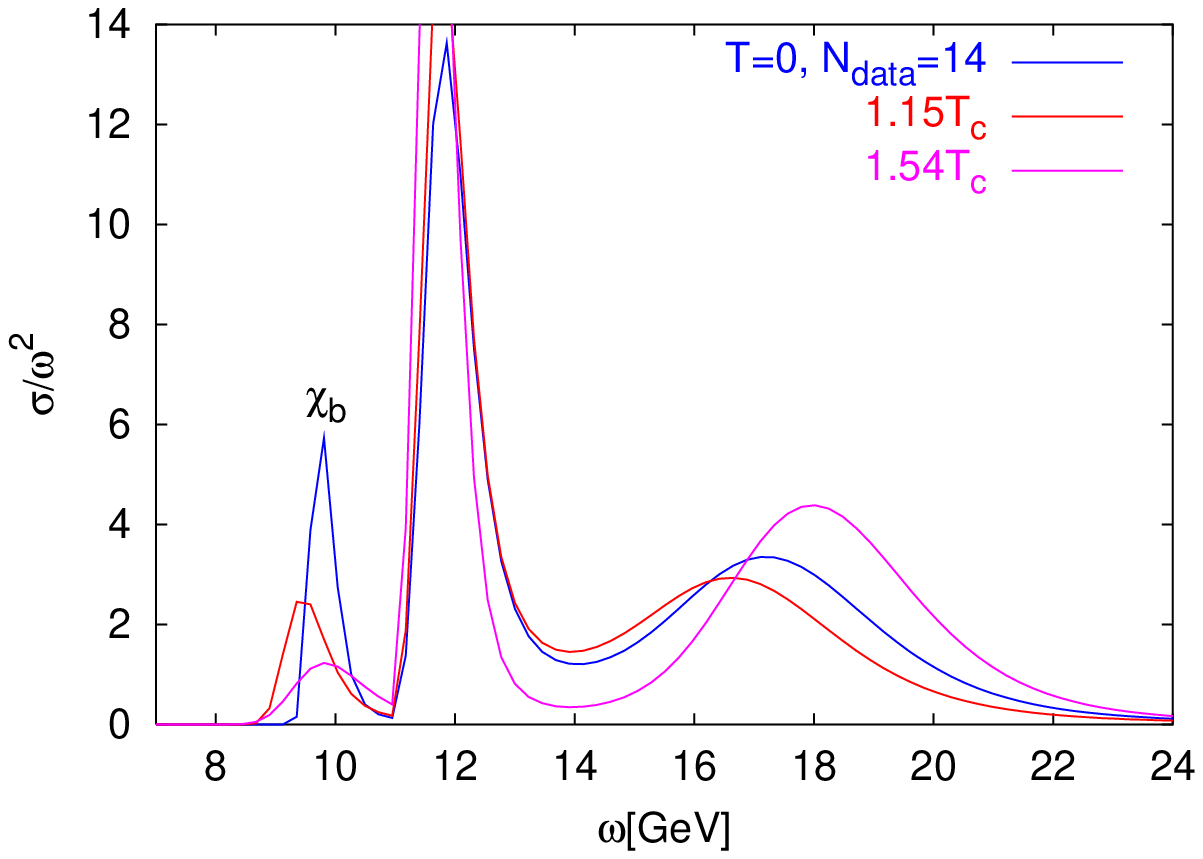}
\caption{Charmonia (left) and bottomonia (right) spectral functions 
in scalar channel for different temperatures.}
\label{scbot}
\end{figure}

With increasing temperature it becomes more and more difficult to
reconstruct the spectral functions, therefore it is useful to study the
temperature dependence of quarkonia correlators first. To separate out the 
trivial temperature dependence due to the integration kernel,
following Ref. \cite{datta04} at each temperature we calculate
the so-called reconstructed correlator  
\begin{equation}
G_{recon}(\tau,T)=\int_{0}^{\infty}d\omega
\sigma(\omega,T=0)K(\tau,\omega,T)
\end{equation}
The ratio of the original and 
the reconstructed correlator should be close to one,
$G(\tau,T)/G_{recon}(\tau,T) \sim 1$, if there is no temperature dependence in
the spectral function. In Fig. \ref{ratioc} we show the ratio of the correlators for charmonium in
pseudoscalar and scalar channels. It is evident that there is little
modification of the correlator in the pseodoscalar channel for temperatures up
to $1.5T_c$. For the scalar channel, however, the change in the correlator is
noticeable at temperatures as low as $1.16T_c$. The results for $\beta=6.1$
and $6.5$ are very similar, thus there is little dependence on the lattice
spacing. 
In Fig. \ref{scbot} we present charmonium and bottomonium spectral functions
for the scalar channel at different temperatures. From these figures the
dissolution of the ground states at temperatures above deconfinement is clear.
For charmonium we tried two different default models $m(\omega)=0.038\omega^2$
and $m(\omega)=1$ and the results are qualitatively the same, which is another
check on systematics.

\section{Summary}
\vspace{-0.8cm}
From our studies of correlators and spectral functions we can conclude that, 
the $1S$ ($\eta_c$, $J/\psi$) charmonium states exist as a resonance in
the deconfined phase at $T\simeq1.5T_c$. On the other hand, the $1P$ ($\chi_{c0}$,
$\chi_{c1}$) charmonium states dissolve at $1.1T_c$. Bottomonium states show similar behavior. 

\section*{Acknowledgments}
\vspace{-0.8cm}
This presentation is done based on work in collaboration with A.~Jakov\'ac,
P.~Petreczky, and K.~Petrov. This work is partially supported by NSF-PHY-0309362.

\end{document}